# High resolution scanning tunneling microscopy and extended X-ray absorption fine structure study of the (5×3) silicide structure on Cu(001)

B. Lalmi[1*], M. Chorro[2], R. Belkhou[1]

[1] HERMES beamline, Synchrotron SOLEIL, L'orme des Merisiers, BP 48, 91192 Saint-Aubin, France

[2] LUCIA beamline, Synchrotron SOLEIL, L'orme des Merisiers, BP 48, 91192 Saint-Aubin, France

**Abstract**

Using low energy electron diffraction (LEED), scanning tunnelling microscopy (STM) and x-ray absorption spectroscopy (XAS) techniques, we have studied the first steps of silicon adsorption onto Cu (001) single crystal substrate. For low coverage (~ 0.5 ML) and after annealing at 100°C, STM images and LEED patterns reveal the formation of an ordered quasi commensurate $(5 \times 3)$ superstructure. From a quantitative analysis of XAS data, we extract the Si-Cu distance and detail the local atomic arrangement of the $(5 \times 3)$ structure.

*Corresponding author: lalmi.boubekeur@gmail.com

**Introduction**

In recent years, various metal-silicon interfaces have been studied, motivated by their relevance in microelectronics applications. Naturally, almost all these studies involve deposition of thin metal films in ultra high-vacuum (UHV) on atomically clean and well-ordered Si substrates [1-8]. Some more recent studies explored the initial stages of the interface formation by investigating the deposition behaviour of silicon on metal, thus many different combinations of metals and Si have been quite extensively studied [8-16]. Silicon/Copper is a typical reactive silicon/metal system, which continues to attract much attention. For low coverage in Si atoms, this leads to the formation of ordered surface alloys and well-defined structures.

Generally, such surface alloys can be formed during growth as well after annealing of the deposited silicon. During the growth of Si onto Cu(110) surface some special features are observed: (i) An ordered surface alloy $c(2\times 2)$-Si/Cu(110) is formed for Si coverage below 0.5ML; (ii) For Si coverage of 0.7 ML, $(2\times 2)$-Si atomic chains considered as a one-dimensional superstructure are observed [11, 17,18,19]. On Cu (111), a surface alloy which exhibits a $(\sqrt{3}\times\sqrt{3})R30°$ structure, was first observed during adsorption and decomposition of silane on the copper surface [20, 21], this superstructure appears at room temperature and still stable at elevated temperatures [22, 23, 24]. On Cu (001) surface, chemical vapor deposition (CVD) is widely used. By decomposition of silane at 150°C, an incommensurate (5×3) structure has been observed [25, 26]. Comparing to the large amount of literature on gas surface interaction for preparations of Si-Cu(001) interfaces, there are only few reports discussing the films prepared by a thermal evaporation method [27, 28].

In the present study, we are interested in the early stages of Si adsorption, thermally evaporated onto Cu (001) surface and in particular, in the formation of a possible ordered overlayer and/or an interface structure which may be induced by thermal annealing. We also aim to elucidate, using three surface sensitive techniques; LEED, STM and EXAFS the precise nature and atomic structure of the copper–silicon interface formed, when an ultrathin (~ 0.5 ML) of silicon is deposited onto Cu (001).

**Experimental details**

All experiments were carried out in ultrahigh-vacuum (UHV) conditions with a base pressure below $2 \times 10^{-10}$ Torr. They were performed in two different experimental equipments. The first apparatus is equipped with the standard tools for surface preparation and characterization: an ion gun for surface cleaning, a low energy electron diffractometer (LEED) for following the structural evolution of the surface, a RT- scanning tunnelling microscopy (STM) for surface characterization at the atomic scale, and an Auger electron spectrometer (AES) for control quality and chemical analysis of the surface.

EXAFS experiments were performed at LUCIA beamline of Synchrotron SOLEIL [29, 30]. The UHV apparatus available at the beamline was mounted downstream the main sample measurement chamber. This apparatus is composed by a sample preparation chamber and a measurement chamber which contains LEED, AES and XAS. X ray absorption spectra were recorded at Si K edge, 1839eV, with an InSb (111) double monochromator and in Total Electron Yield mode (TEY). The Cu (001) surface was probed at normal incidence with a polarization vector *E* parallel to the surface and lying in the horizontal plane. The size of the beam was about 1mm$^2$. A sample of silicon crystal was measured and used as reference for analysis and beamline energy calibration. Data treatments of the raw spectra, i.e., energy calibration, normalization, and EXAFS spectra extraction, were performed with ATHENA program. Analyses of the EXAFS spectra and fits were performed with ARTEMIS program. These programs are included in the *IFEFFIT* package for XAS analysis [31].

Before any experiment, whatever technique used, the Cu (001) surface was cleaned by cycles of Ar$^+$ ion bombardment ($5 \times 10^{-5}$ Torr at 500 eV) followed by annealing at 550°C. The sample temperature was evaluated from a thermocouple spot-welded on the sample holder very close to the crystal. The sputtering-annealing cycles were performed until a sharp $(1 \times 1)$ LEED pattern was observed. Silicon deposition was carried out by thermal evaporation from a silicon wafer heated by Joule effect. The silicon evaporation rate was calibrated by AES and a quartz microbalance [9, 30].

**Results and discussions**

After the deposition of a sub-monolayer (~0.5 ML) of silicon onto Cu(001) followed by annealing at 100°C during 10 minutes, a new superstructure was observed by LEED as shown in figure 1. The LEED pattern displays several spots which appear bright and intense, indicating that the surface layer presents a high crystalline quality. This LEED pattern was simulated under "LEED pattern 3.0" software and can be interpreted as the superposition of orthogonal domains forming an incommensurate $(5\times3)$ superstructure. Indeed spots from the superstructure are not strictly aligned according the [10], [01] directions and some other spots appear also doubled. A good agreement between the experimental LEED pattern in Figure 1(a) and the simulated one in Figure 1(b) is obtained by considering lattice parameters of the superstructure unit cell $a_s$= 1.292 nm and $b_s$= 0.746 nm. The unit cells corresponding to the substrate and the superstructure in reciprocal space are indicated in the Figure 1(b).

To better understand the nature and structural properties of this reconstruction, STM images of the surface in real space were performed. Figure 2(a) presents a large area topographic scan of the copper-silicon surface indicating the size and degree of the copper-silicon domains that can be formed. The size of the topographic scan is of approximately $42\times42$ nm$^2$, it displays two perpendicular domains and a long-range corrugation. These domains are formed by periodic stripes oriented along the [100] and [010] crystallographic directions of Cu(001) substrate. The stripes observed in each domain are ~ 1.29 nm apart (~ $5a$ where $a$ = 0.256 nm is the copper substrate lattice spacing) and have a corrugation of 0.03± 0.01 nm as shown by the line scan reported in Fig. 2(b). The $5a$ stripes reflect the modulation of the surface layer with respect to the substrate.

Figure 2(a) also displays an abrupt termination of the striped structure in the domains boundary. The rows observed at the edge of each domain are apparently unaffected by the proximity of the adjacent domain. This indicates that the stresses at the domains borders can be weak. In the case of relatively strong constrains, elastic (distortion in the periodic arrangement) and/or plastic (nucleation of dislocations) relaxations would have been induced and observed in our STM images.

Overall, in all STM images taken over several areas of the surface, the reconstructed two-dimensional copper-silicide layer seems to be homogenous and continuous over a large-scale of the surface. Only few defects randomly distributed in some locations of the surface were observed. These defects are possibly due to spare copper released from the surface during the formation of the overlayer or more likely, copper clusters deposited from the tip.

Figure 3(a) presents a high resolution topograph of the surface recorded on the 5a striped structure oriented along the [010] direction. As can be seen, a new periodicity running approximately perpendicular to the 5a stripes is present. This periodicity is of about 0.75 nm which is approximately equal to 3a. STM images clearly confirm that the copper-silicide has a nearly (5×3) structure, already observed in the LEED patterns. The (5×3) unit cell is evident, but in addition, it is possible to observe more detail within that unit cell. Indeed, protrusions with high contrast forming an alternate zigzag are clearly discernable. Since scanning tunnelling microscopy probe the filled and unfilled electronic states of the surface depending on the sample bias applied during the experiment, the protrusions observed within the (5×3) unit cell could be assigned to the atomic species having a relatively high electronic density in the silicide layer.

In the present study, the surface layer contains two different atomic species that have different electronic properties, so care must be taken in the interpretation of the STM topographs.

The weak amplitude of corrugation observed for both periodicities (5*a* and 3*a)* can be attributed to a difference in electronic proprieties of the two elements forming the silicide layer. Such low corrugations cannot arise from silicon atoms sitting on the top of the copper surface, but must originate with silicon atoms incorporated into the surface substrate, as is generally observed in similar systems [9, 32, 33].

The protrusions observed in Fig.3(a) represent local features of high electronic density, they probably correspond to silicon atoms, if we consider the electronegativity of silicon is stronger than the one of copper in the silicide layer ( according to the EXAFS results see below).

With the sole use of STM images, it is extremely difficult to resolve atomically the (5×3) structure, in order to extract more information and give local details about the atomic arrangement of each element silicon/copper in the silicide layer, EXAFS measurements were performed.

Figure 4 presents the XANES part of the raw absorption spectrum of monolayer silicide recorded in TEY. The edge jump is 0.015 (a.u.) and corresponds to an amount of silicon of about 0.5ML. Figure 5 presents the comparison of the normalized spectra of silicon crystal and silicide monolayer. The edge energy is defined at the inflexion point for both spectrum and set to 1839eV for the silicon crystal used as reference sample. One can observe an edge shift of the silicide spectrum of -0.8eV. This downshift can be related to the negative charge density on the silicon atom due to the difference in electronegativity between silicon and copper atoms in the copper silicide. Figure 6 presents the EXAFS spectrum of copper silicide and its Fourier transform. The spectrum shows well defined oscillations up to k=9.5Å$^{-1}$. The Fourier transform reveals the first coordination shell distance at about 2.2Å (uncorrected from phase shifts) and a very small second contribution at 3.5Å. Analysis of the silicon crystal EXAFS spectra was done in order to determine the amplitude reduction factor of silicon, $S_0^2 = 0.9$ (see Table 1). This factor was used as reference to fit the silicide EXAFS spectra. The amplitude of the EXAFS oscillations is proportional to the number of atoms in a considered coordination shell. As we do not know exactly the local atomic structure we first evaluated the approximate number of copper neighbors by fitting the silicide spectra with a simple structural model consisting on a single Si-Cu bond. Amplitude and phase shifts of the signal are theoretically determined with the FEFF 6 code [31]. Approximately 10 copper atoms were found in the first coordination shell. As the structure is a monolayer copper-silicide we tried several models to account for the 10 copper atoms. The measurement being in normal incidence and for a polarization vector *E* lying in the horizontal plane for the incoming photons we also had to consider the polarization as follow.

For non-centro-symmetric systems, the Exafs model has to consider the polarization effect according to the equation (1) [34]:

$$X_{i,j}(k) = -\frac{3}{kR_i^2} S_0^2 \cos^2(\theta_i) \times A(k,\pi) \sin(2kR_i + 2\delta l + \Phi) \quad (1)$$

This equation represents the contribution to the EXAFS signal from one atom $i$ at position $R_i$, located in a direction $\theta_i$ with respect to the polarization vector $E$, in the j$^{th}$ shell (see figure 7). Then one has to sum the contributions for all the atoms of the considered j$^{th}$ shell, where a 'shell' stands for identical atoms (same phase shifts functions) at the same distance (same $R$).

$$X_j(k) = \sum_i X_{i,j}(k) \quad (2)$$

Finally one obtains the equation to the EXAFS signal of the j$^{th}$ shell at distance R from the absorbing atoms, where the apparent coordination number $N_j$ appears

$$X_j(k) = -\frac{1}{kR^2} S_0^2 N_j \times A(k,\pi) \sin(2kR + 2\delta l + \Phi) \quad (3)$$

Therefore the apparent coordination number $N_j$ can be different from the total number of atoms in the shell. The only valid model is one silicon atom surrounded by 6 copper atoms placed at the vertices of a hexagon as presented in Figure 7a. In this configuration, the first shell contains 6 atoms and the apparent coordination number is equal to $N = 9$. This result remains unchanged by rotating the orientation of the hexagon (i.e. of the monolayer) with respect to the polarization vector. So the EXAFS signals coming from the two perpendicular domains observed in STM are identical. The EXAFS signal generated by this structural model was determined with FEFF 6 code including the polarization effect. Results of the fit are presented in Table 1. The Si-Cu distance in the copper silicide is 2.48±0.02Å. This value is consistent with the one measured using EXAFS for a silicide monolayer prepared by silane thermal decomposition on Cu (111) [22]. However the distance is a bit larger than the 2.46Å presented by *Graham et al.* for Si on Cu (001) [25]. But these authors used experimental techniques (scanning tunneling microscopy and helium atoms scattering) much less accurate in distance determination than EXAFS. The very small signal present at about 3.5 Å cannot be attributed to the silicon atom of the second coordination shell located around 4.3 Å. Actually *Graham et al.* in ref [25] invoke a shearing of the hexagonal Cu-Si plane, which will induce a spreading of the mid-range distances and explain that the Si-Si second neighbors cannot be seen here. So the contribution at 3.5 Å probably arises from the Cu atoms of the copper surface underneath, and its small amplitude means that this surface is incommensurate with the silicide one; the EXAFS signal is therefore killed by disorder effects.

Finally, according to the nearly $(5\times 3)$ periodicity observed in STM, and to the local atomic environment detailed by EXAFS, we can detail the structure of the silicide overlayer as illustrated in figure 7b. One can see that silicon atoms at the vertices of the rectangular unit cell are nearly on top of substrate copper atoms. The unit cell is quasi commensurate with the substrate in the 5a direction. Indeed, the distance between the silicon atoms at the extremities of the superstructure is of 12.94Å that is 0.14Å larger than the 5a periodicity (12.8Å). However it is not commensurate in the 3a direction. In the 3a direction the superstructure unit cell is smaller than what we should have for the 3a periodicity, 7.47Å instead of 7.68Å. From the structure of the unit cell, the stoichiometry of the monolayer can be determined as $Cu_2Si$.

**Conclusion**

In summary, adsorption of a sub-monolayer of thermally evaporated silicon onto Cu(001), was investigated by three uniquely sensitive techniques LEED, STM and EXAFS. The silicon reacts with the surface copper atoms to form a two-dimensional copper-silicide. LEED patterns and STM images show that the surface alloy displays two large perpendicular domains with a $(5\times 3)$ reconstruction. XAS measurements lead to a determination of a Si-Cu distance of 2.48±0.02Å. According to the STM observations and quantitative analysis of EXAFS data, we propose that the surface alloy formed, has an incommensurate $(5\times 3)$ structure with $Cu_2Si$ stoichiometry.

**References**

1. K. Kataoka, K. Hattori, Y. Miyatake, and H. Daimon, Phys. Rev. B **74**, 155406 (2006).

2. D. Grozea, E. Bengu, L.D. Marks, Surf. Sci. **461**, 23 (2000).

3. Z.Q. Zou, D. Wang, J. J. Sun, and J. M. Liang, J. Appl. Phys. **107**, 014302 (2010)

4. P. Höpfner, M. Wisniewski, F. Sandrock, J. Schäfer, and R. Claessen, Phys. Rev. B **82**, 075431 (2010).

5. L. Casalis, A. Citti, R. Rosei, and M. Kiskinova, Phys. Rev. B **51**, 1954 (1995).

6. P. A. Bennett and H von K¨anel, J. Phys. D: Appl. Phys. **32**, 71 (1999).

7. D. Bolmont, V. Mercier, P. Chen, H. Lüth, C.A. Sébenne, Surf. Sci. **126**, 509 (1983).

8. R. J. Phaneuf, Y. Hong, S. Horch, and P. A. Bennett, Phys. Rev. Lett. **78**, 4605 (1997).

9. B. Lalmi, C.Girardeaux, A. Portavoce, C. Ottaviani, B.Aufray, J. Bernardini, Phys. Rev. B **85**, 245306 (2012).

10. B. Lalmi, H. Oughaddou, H. Enriquez, A.Kara, S.Vizzini, B. Ealet, B. Aufray, Appl. Phys. Lett. **97**, 223109 (2010).

11. J.A. Martin-Gago, C. Rojas, C. Polop, J. L. Sacedon, E. Roman, A. Goldoni, G. Paolucci, Phys. Rev. B **59**, 3070 (1999).

12. R. dudde, H. Bernhoff, B. Reihl, Phys. Rev. B **41**, 12029 (1990).

13. Y. Wang, V. Bykov, J.W. Rabalais , Surf. Sci. **319**, 329-336 (1994).

14. Yi. Wang and S. J. Sibener, J.Phys.Chem.B **106**, 12856 (2002).

15. H. Enriquez, A. Mayne, A. Kara, S. Vizzini, A. Roth, B. Lalmi, A. Seitonen, B. Aufray, T. Gerber, R. Belkhou, G. Dujardin, H. Oughaddou , Appl. Phys. Lett, **101**, 021605 (2012).

16. A. Portavoce, B. Lalmi, G. Tréglia, C. Girardeux, D. Mangelinck, B. Aufray, J. Bernardini, Appl. Phys. Lett, **95,** 023111, (2009).

17. C. Polop, J. L. Sacedon, J. A. Martin-Gaogo, Surf. Sci. **402-404**, 245-248 (1998).

18. C. Polop, C. Rojas, J. A. Martin-Gago, R. Fasel, J. Hayoz, D. Naumovi and P. Aebi, Phys. Rev. B **63**, 115414 (2001).

19. C. Rojas, F. J. Palmares, M. F. Lopez, A. Goldoni, G. Paolucci and J. A. Martin-Gago, Surf. Sci **778**, 454-456 (2000).

20. E. M. McCash, M. A. Chesters, P. Gardner, S. F. Parker, Surf. Sci, **225**, 273-280 (1990).

21. A. W. Robinson, P. Gardner, A. P. J. Stampfl, R. Martin, G. Nyberg, Surf. Sci, **387**, 243-256, (1997).

22. T. Kanazawa, Y. Kitajima, T. Yokoyama, S. Yagi, A. Imanishi, T. Ohta, Surf. Sci. **357-358**, 160-164 (1996).

23. H. Menard, A.B. Hom, S.P. Tear, Surf. Sci. **47**, 585 (2005).

24. J. S. Tsay, A. B. Yang, C. N. Wu, F. S. Shiu, Thin solid films, **515**, 8285-8289 (2007).

25. A.P. Graham, B. J. Hinch, G.P. Kochanski, E. M. MacCash, W. Allison, Phys. Rev. B **50**, 15304 (1994).

26. L. H. Dubois, R. G. Nuzzo, *Langmuir* **1**, 663 (1985).

27. B. Lalmi, C. Girardeaux, A. Portavoce, J. Bernardini, B. Aufray, J. Nanosci. Nanotechnol. **9**, 4311 (2009).

28. B. Lalmi, C. Girardeaux, A. Portavoce, J. Bernardini, B. Aufray, Defect and diffusion forum. **289-292**, 601-606 (2009).

29. R. Gunnella, JY. Veuillen, Tan. TAN, A-M. Flank, Phys. Rev. B **57**, 4154 (1998).

30. A-M. Flank, G. Cauchon, P. Lagarde, S. Bac, M. Janousch, R. Wetter, J.-M. Dubuisson, M. Idir, F. Langlois, T. Moreno, D. Vantelon, Nuclear Instruments and Methods in Physics Research Section B: Beam Interactions with Materials and Atoms, **246**, 269–274 (2006).

31. M. Newville, IFEFFIT, http://cars9.uchicago.edu/ifeffit (2009).

32. B. Lalmi, Phd thesis, Université Paul Cézanne-Marseille III (2009).

33. J. A. Martín-Gago, R. Fasel, J. Hayoz, R.G. Agostino, D. Naumovi, P. Aebi and L. Schlapbach, Phys. Rev. B **55,** 12896 (1997).

34. D. C. Koningsberger and R. Prins, X-Ray absorption, Principles, applications, Techniques of EXAFS, SEXAFS and XANES, ISBN 0-471-87547-3 (New York : John Wiley) (1988).

**Figure captions**

**Figure 1:**

a) LEED pattern corresponding to the quasi $(5\times 3)$ superstructure obtained after the deposition of ~ 0.5 Si ML on Cu(001) (Ep = 62eV), b) A simulation of the LEED pattern corresponding to the superstructure with two perpendiculars domains.

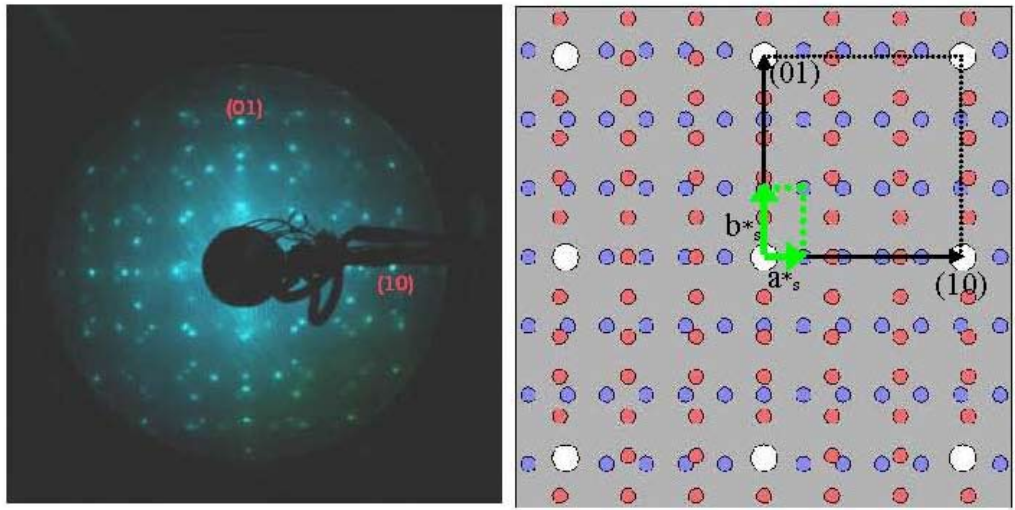

**Figure 2:**

a) A topograph of the surface alloy with two perpendicular large domains (Imaging conditions: 1.1 V Sample bias and 0.9 nA tunnel current), b) line scan.

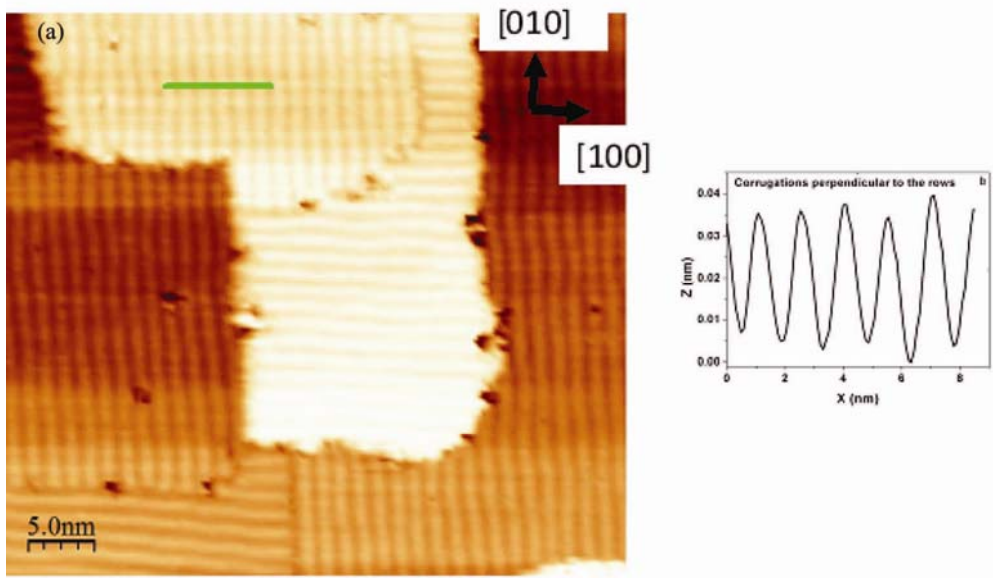

**Figure 3:**

a) Atomically resolved filled-states STM image corresponding to the (5×3) superstructure showing detail within the unit cell reconstruction (V = 0.80 V, I = 1.2 nA), the unit cell is indicated, b) line scan.

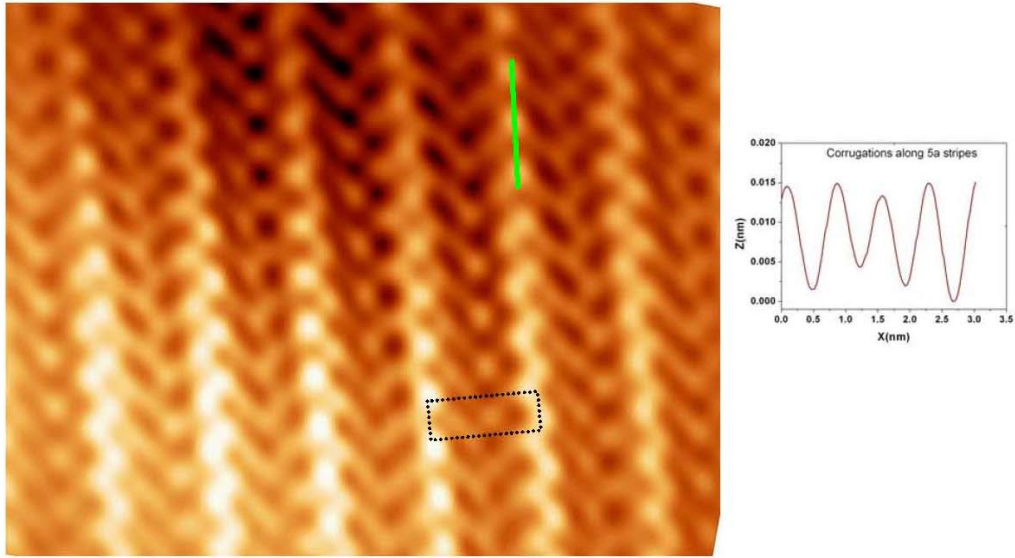

**Figure 4:**

Absorption spectrum of the silicide monolayer.

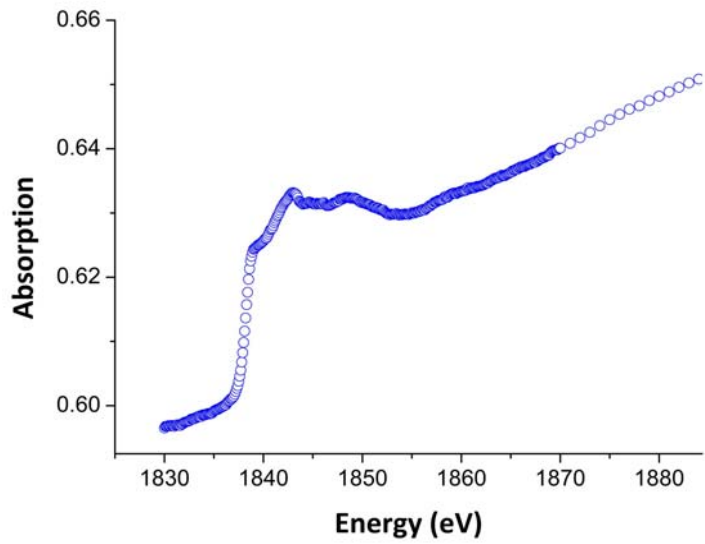

**Figure 5:**

Left: Normalized spectra of silicon crystal (black line open square) and of copper silicide (red line open circle). Right: Normalized XANES region. The silicide monolayer shows a down shift of -0.8eV.

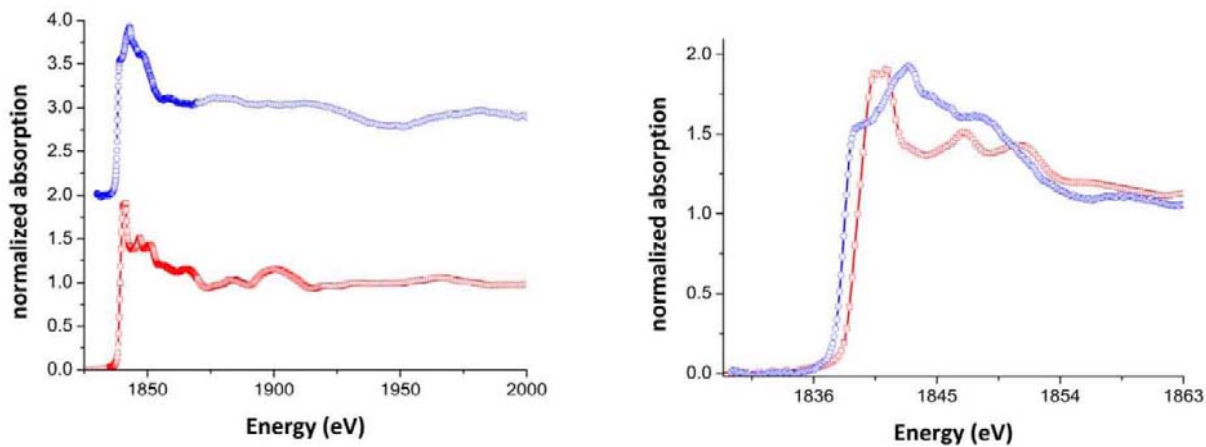

**Figure 6:**

Left: EXAFS spectra (open circle) and best fit (red solid line). Right: Fourier Transform (open circle) and best fit for Cu$_2$Si stoichiometry (red solid line).

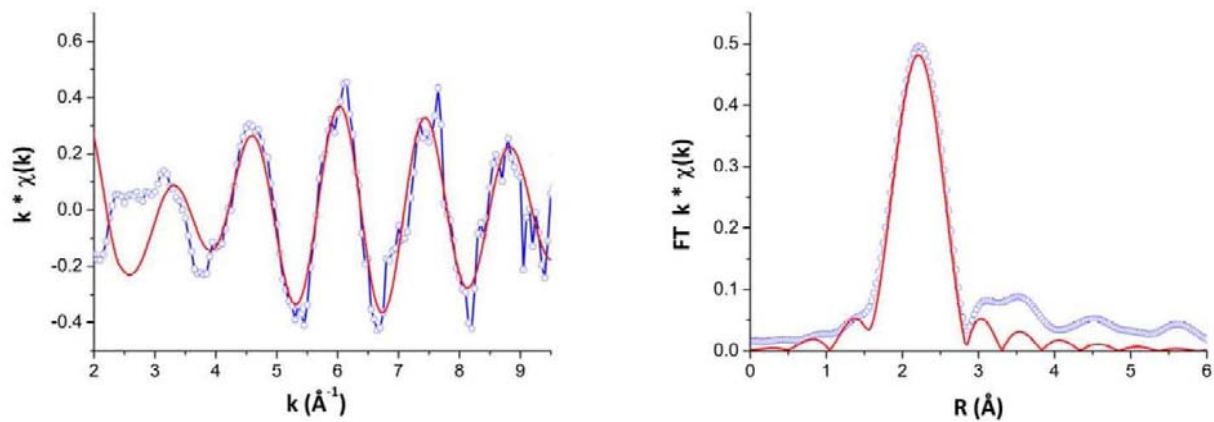

**Figure 7:**

a): Schematic representation of the atomic environment of silicon atoms in the silicide monolayer (blue silicon atoms, orange copper atoms). The central silicon atom stands for the absorbing atom. b): Schematic representation of the monolayer silicide onto the Cu (001) surface. The links between the monolayer atoms are guideline for the hexagonal array.

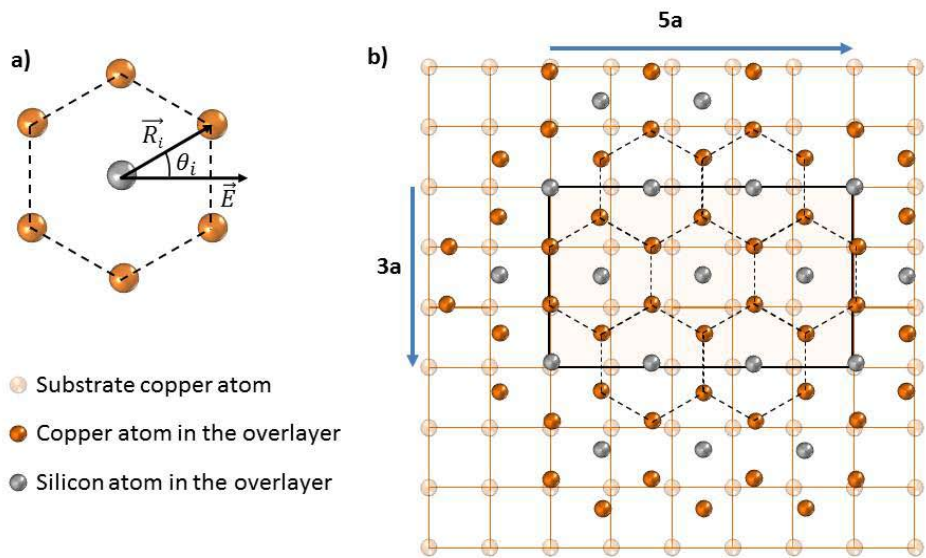

**Table captions**
**Table 1:**

Results of first shell fit of silicon crystal and copper-silicide monolayer samples.

| Sample | structure | ΔE (eV) | R (Å) | $\sigma^2$ (x$10^{-3}$ Å$^2$) | $s_0^2$ | N |
|---|---|---|---|---|---|---|
| Silicon | Fd-3m | 5.6±1.2 | 2.355±0.003 | 4±0.1 | 0.9 | 4 |
| Silicide | model | 4.6±2.5 | 2.48±0.02 | 6±1 | 0.9 | 9 |